\documentclass[twocolumn,superscriptaddress,preprintnumbers,amsmath,amssymb,prl]{revtex4}
\UseRawInputEncoding
\usepackage{graphicx}

\begin{document}
\title{Theory-experiment comparison for the Casimir force between metallic
test bodies: A spatially nonlocal dielectric response}

\author{
G.~L.~Klimchitskaya}
\affiliation{Central Astronomical Observatory at Pulkovo of the
Russian Academy of Sciences, Saint Petersburg,
196140, Russia}
\affiliation{Peter the Great Saint Petersburg
Polytechnic University, Saint Petersburg, 195251, Russia}

\author{
V.~M.~Mostepanenko}
\affiliation{Central Astronomical Observatory at Pulkovo of the
Russian Academy of Sciences, Saint Petersburg,
196140, Russia}
\affiliation{Peter the Great Saint Petersburg
Polytechnic University, Saint Petersburg, 195251, Russia}
\affiliation{Kazan Federal University, Kazan, 420008, Russia}

\begin{abstract}
It has been known that the Lifshitz theory of the Casimir force comes
into conflict with the measurement data if the response of conduction
electrons in metals to electromagnetic fluctuations is described by
the well tested dissipative Drude model. The same theory is in a very
good agreement with measurements of the Casimir force from graphene
whose  spatially nonlocal electromagnetic response is derived from
the first principles of quantum electrodynamics. Here, we propose the
spatially nonlocal phenomenological dielectric functions of metals
which lead to nearly the same response, as the standard Drude model,
to the propagating waves, but to a different response in the case
of evanescent waves. Unlike some previous suggestions of this type,
the response functions considered here depend on all components of
the wave vector as is most natural in the formalism of specular
reflection used. It is shown that these response functions satisfy
the Kramers-Kronig relations. We derive respective expressions for
the surface impedances and reflection coefficients. The obtained
results are used to compute the effective Casimir pressure between
two parallel plates, the Casimir force between a sphere and a plate,
and its gradient in configurations of the most precise experiments
performed with both nonmagnetic (Au) and magnetic (Ni) test bodies.
It is shown that in all cases (Au-Au, Au-Ni, and Ni-Ni test bodies)
the predictions of the Lifshitz theory found by using the dissipative
nonlocal response functions are in as good agreement with the
measurement data, as was reached previously with the dissipationless
plasma model. Possible developments and applications of these
results are discussed.
\end{abstract}

\maketitle
\newcommand{\kb}{{k_{\bot}}}
\newcommand{\skb}{{k_{\bot}^2}}
\newcommand{\vk}{{\mbox{\boldmath$k$}}}
\newcommand{\rv}{{\mbox{\boldmath$r$}}}
\newcommand{\ve}{{\varepsilon}}
\newcommand{\Te}{{\varepsilon^{\rm Tr}}}
\newcommand{\Le}{{\varepsilon^{\,\rm L}}}
\newcommand{\Tle}{{\varepsilon_l^{\rm Tr}}}
\newcommand{\Lle}{{\varepsilon_l^{\,\rm L}}}
\newcommand{\Tne}{{\varepsilon_n^{\rm Tr}}}
\newcommand{\Lne}{{\varepsilon_n^{\,\rm L}}}
\newcommand{\zo}{{(z;\omega,k_x)}}
\newcommand{\oz}{{(\omega,k_x,k_z)}}
\newcommand{\xk}{{(i\xi_l,k_{\bot})}}
\newcommand{\VT}{{v^{\rm Tr}}}
\newcommand{\VL}{{v^{\,\rm L}}}
\newcommand{\vTn}{{v_n^{\rm Tr}}}
\newcommand{\vLn}{{v_n^{\,\rm L}}}
\newcommand{\rMn}{{r_{\rm TM}^{(n)}}}
\newcommand{\rEn}{{r_{\rm TE}^{(n)}}}
\newcommand{\zMn}{{Z_{\rm TM}^{(n)}}}
\newcommand{\zEn}{{Z_{\rm TE}^{(n)}}}


\section{Introduction}

As predicted by Casimir \cite{1}, two parallel uncharged ideal metal
planes separated by a distance $a$ at zero temperature should attract
each other by the force
\begin{equation}
F(a)=-\frac{\pi^2}{240}\,\frac{\hbar c}{a^4},
\label{eq0}
\end{equation}
\noindent
which depends on the Planck constant $\hbar$ and the speed of light $c$.
According to Casimir, this force is caused by the zero-point oscillations of
quantized electromagnetic field whose spectrum is altered by the presence
of ideal metal planes. Within the formalism of quantum electrodynamics,
in the absence of planes the zero-point energy is given by an integral over
a continuous wave vector $\vk$. In the presence of planes, however, the tangential
component of electric field vanishes on the plane surfaces and the component
$k_z$ in direction perpendicular to the planes becomes discrete.
The modified zero-point energy of the electromagnetic field is an integral
over $k_x$ and $k_y$, but a discrete sum in $k_z$. Although both zero-point
energies in the absence and in the presence of planes are infinitely large,
their difference is finite. As a result, the negative derivative of this
difference with respect to $a$ is equal to the Casimir force (\ref{eq0}).

More recently, it was understood that the Casimir force, as well as the
more familiar van der Waals force, belongs to a wide class of physical
phenomena determined by the zero-point and thermal fluctuations of the
electromagnetic field. Lifshitz \cite{2} created the general theory of
forces of this kind acting between two parallel material plates
(semispaces) kept at any temperature in thermal equilibrium with the
environment. In the framework of the Lifshitz theory, the ideal metal
boundary conditions are replaced by the electrodynamic continuity
conditions which take into account real material properties by means of
the frequency-dependent dielectric permittivity and (for magnetic plates)
magnetic permeability.

There are two main approaches to a derivation of basic expressions of the
Lifshitz theory for the free energy and force of fluctuation origin.
One of them is based on the fluctuation-dissipation theorem of statistical
physics \cite{2,3} and another one, which goes back to Casimir, on quantum
field theory with appropriate boundary conditions \cite{4,5,6,7}.
By the frequently used present-day terminology, the name van der Waals force
refers to the fluctuation-induced forces at separations of a few nanometers
which do not depend on $c$. The term Casimir force is used in all remaining
cases, i.e., when the interaction of fluctuation origin depends on $\hbar$,
$c$, material properties and temperature. By now the Lifshitz theory is
generalized for the case of boundary surfaces of arbitrary geometric shape
\cite{8,9,10,11,12}.

A systematic investigation of the thermal Casimir force between parallel
plates made of real metals traces back to 2000. In Ref.~\cite{13} it was
shown that if the low-frequency dielectric response of metals is described
by the realistic Drude model taking a proper account of the relaxation
properties of conduction electrons, the Lifshitz theory predicts large
thermal correction which arises at short separations of a few hundred
nanometers at room temperature and decreases the force magnitude.
The thermal correction of this type does not arise if the low-frequency
dielectric response of metals is described by the plasma model which
disregards the relaxation properties of conduction electrons \cite{14}.
It should be remembered, however, that the plasma model is in fact
applicable only at high frequencies in the range of infrared optics
where the relaxation properties do not play any role.
The surprising thing is that the Casimir entropy calculated with the
Lifshitz theory employing the Drude model violates the third law of
thermodynamics (the Nernst heat theorem) for metals with perfect crystal
lattices \cite{15,16,17,18,19,20,21}
(the Nernst heat theorem is followed for metals with the defects of structure
\cite{22,23,24}, but this does not solve the problem because perfect
crystal lattice is a system with the nondegenerate ground state, so that it
should satisfy the third law of thermodynamics). If, however, the Casimir
entropy is found using the plasma model dielectric response, the Nernst
heat theorem is satisfied without trouble \cite{15,16,17,18,19,20,21}.

Of even greater surprise is that the theoretical predictions of the Lifshitz
theory using the Drude dielectric response at low frequencies have been excluded
by the measurement data of many experiments performed with both nonmagnetic (Au)
and magnetic (Ni) metallic test bodies by the two different experimental groups
\cite{25,26,27,28,29,30,31,32,33,34,35,36,37}.
The same measurement data were found to be in a very good agreement with the
predictions of the Lifshitz theory employing the plasma dielectric response
\cite{25,26,27,28,29,30,31,32,33,34,35,36,37}.
In the most striking experiment of Ref.~\cite{33} using the differential
measurement scheme, a difference between the excluded and confirmed theoretical
predictions was by up to a factor of 1000. Thus,
the Drude mode, which provides an adequate description of numerous optical and
electrical physical phenomena, does not work in application
to the Casimir force.

Note that the measurement data of one experiment performed at separations of a
few micrometers were found to be in better agreement with theoretical predictions
using the Drude model \cite{38}. To obtain this conclusion, the Casimir force
calculated using the Drude or the plasma model was subtracted from by the order
of magnitude larger measured force. The obtained differences were fitted to the
theoretical electric force originating from the surface patches. The results of Ref.~\cite{38}
were shown to be, however,  uncertain because in this experiment the surface
imperfections, which are unavoidably present on a surface of the used spherical
lens of centimeter-size radius, were ignored \cite{39}. Recent Casimir experiment
performed in the micrometer separation range, where the patch potentials were
directly measured by means of Kelvin probe  microscopy, demonstrated an agreement
with theoretical predictions using the plasma model and excluded those using the
Drude model \cite{37}.

An apparent disagreement of theoretical predictions of the fundamental Lifshitz
theory using the Drude model, which was fully validated in the area of
electromagnetic and optical phenomena other than the Casimir effect, with the
measurement data, as well as with the basic principles of thermodynamics, is
puzzling and calls for a satisfactory explanation. This subject was hotly
debated in the literature starting from 2000 and many attempts have been made
directed to reaching an agreement with thermodynamics, looking for some
unaccounted systematic effects in measurements of the Casimir force, or
developing the more exact theory for a sphere-plate geometry used in experiments.
Considerable advances have been made in this way (see Refs.~\cite{40,41,42,43,44,45}
for a review), but the ultimate resolution of the Casimir puzzle still remains to
be found.

One possible approach to understanding of this problem is that
all models of the electromagnetic response of macroscopic bodies are
in some sense phenomenological, either relying on assumptions about
bulk properties or upon very simplified models of the response of
individual atoms to an applied field. This means that no model
derived in any macroscopic framework  using, e.g., the Boltzmann transport
equations or Kubo theory can be expected to work under
all circumstances, especially, in the very extreme conditions in
Casimir force experiments. From this standpoint, it
is not at all surprising that some
physical situations give results in contrast with the predictions of
commonly used models. There are, however, some limitations following
from fundamental physics that exacerbate the situation. Thus, according to
Maxwell equations, the dielectric response of metals to electromagnetic field
in the quasistatic limit is inverse proportional to the frequency.
The Drude model satisfies this demand and describes the relaxation properties of
conduction electrons whereas the dielectric permittivity of the plasma model
is inverse proportional to the second power of frequency and does not
describe the relaxation properties.

At one time it was hoped that a resolution of this problem may come from the
investigation of graphene which is a two-dimensional sheet of carbon atoms packed
in the hexagonal crystal lattice. At energies below a few eV, which are
characteristic for the Casimir effect at not too short separations, graphene is
well described by the Dirac model as a set of massless or very light electronic
quasiparticles satisfying the Dirac equation where $c$ is replaced with the
Fermi velocity $v_{\,\rm F}$ \cite{46,47}.
The spatially nonlocal dielectric response of graphene to electromagnetic field
was found precisely on the basis of first principles of quantum electrodynamics
using the polarization tensor \cite{48,49,50,51}.
The Lifshitz theory employing this dielectric response turned out to be in
agreement with experiments on measuring the Casimir force in graphene systems
\cite{52,53,54,55} and with the Nernst heat theorem \cite{56,57,58,59,59a}.
The question arises of whether the spatially nonlocal dielectric response could
be helpful for a resolution of the Casimir puzzle.

Unfortunately, an application of the conventional spatially nonlocal dielectric
permittivities derived in the literature for theoretical description of the
anomalous skin effect, screening effects etc. \cite{60,61,62,63,64,65,66a}, although
leaves room for a resolution of thermodynamic problems, does not remedy a
contradiction between the Lifshitz theory and the measurement data
\cite{66,67,68,69,70}. In this situation, some phenomenological models are
worth consideration.

With this approach, Ref.~\cite{71} proposed the spatially nonlocal dielectric
permittivities which describe nearly the same response, as the Drude model,
to electromagnetic waves on the mass shell (i.e., to the propagating waves),
but quite a different response, than the Drude model, to the off-the-mass-shell
waves, which are also called evanescent. These permittivities depend only on
the magnitude of the wave vector projection on the plane of Casimir plates
$\kb$. It was shown that they satisfy the Kramers-Kronig relations \cite{71},
and the respective Casimir entropy goes to zero with vanishing temperature in
agreement with the Nernst heat theorem \cite{72}. With the aim of solving the above
problems, spatially nonlocal permittivities were also introduced in Ref.~\cite{73a}.

The most important thing is that the proposed nonlocal permittivities bring the
Lifshitz theory in agreement with the measurement data of experiments performed
with two nonmagnetic (Au) \cite{71} and two magnetic (Ni) \cite{73} test bodies.
To perform the theory-experiment comparison, the expressions for the reflection
coefficients entering the Lifshitz formula via the surface impedances were used.
The latter have been found in Refs.~\cite{61,62} for nonmagnetic metals in the
approximation of specular reflection of electrons on the boundary surfaces as
some functionals of nonlocal dielectric permittivities. For the case of magnetic
metals, similar surface impedances were recently derived in Ref.~\cite{73}.

In this paper, we suggest the spatially nonlocal response functions which,
similar to Ref.~\cite{71}, lead to the same results, as the standard Drude model,
for electromagnetic waves on the mass shell, but to alternative results for the
off-the-mass-shell fields. However, unlike Ref.~\cite{71}, the response functions
considered below depend on all components of the wave vector $\vk$ what is
fully consistent with
the used formalism of surface impedances in the approximation of
specular reflection developed in Refs.~\cite{61,62}  (see Sec.~II for more
detail).

We calculate the surface impedances and reflection coefficients for two independent
polarizations of the electromagnetic field in the case of Au and Ni surfaces.
The obtained reflection coefficients for Au surfaces are used to perform
computations of the effective Casimir pressure between two Au-coated plates \cite{27}
and the Casimir force between an Au-coated sphere and an Au-coated plate \cite{37}
in the experiments performed by means of a micromechanical torsional oscillator.
Using the same reflection coefficients, we also compute the gradient of the Casimir
force between an Au-coated sphere and an Au-coated plate in the experiments performed
in different separation regions by means of an atomic force microscope \cite{29,36}.
With the help of reflection coefficients on both Au and
Ni surfaces, we perform computations of
the gradient of the Casimir
force between an Au-coated sphere and a Ni-coated plate
measured in the experiment \cite{30}.
Finally,  the gradient of the Casimir
force between a Ni-coated sphere and a Ni-coated plate  is
computed which was measured in Refs.~\cite{31,32}.
According to our results, in all cases the predictions of the Lifshitz theory
using the suggested nonlocal response functions, which take the proper account
of the relaxation properties of conduction electrons in the region of propagating
waves, are in a very good agreement with the measurement data. Future applications
of these results are discussed.

The paper is organized as follows. In Sec.~II, we present the formalism of the
Lifshitz theory in the approximation of specular reflection. In Sec.~III, the
spatially nonlocal response functions depending on all components of the wave
vector are introduced and their properties are investigated.
Section~IV is devoted to computations of the effective Casimir pressure and
Casimir force in different experiments using a micromechanical torsional
oscillator. In Secs.~V and VI computations of the gradient of the Casimir force
are performed between nonmagnetic and with magnetic test bodies, respectively,
measured by means of an atomic force microscope. In Sec.~VII, the reader will
find our conclusions and a discussion.

\section{Formalism of the Lifshitz theory in the approximation of specular
reflection}

The Casimir free energy per unit area and pressure in the configuration of
two thick parallel plates (semispaces) spaced at separation $a$ at temperature
$T$ in thermal equilibrium with the environment are given by the famous Lifshitz
formulas \cite{2,3} (see also \cite{40,41} for modern notations in terms of
the reflection coefficients used below)
\begin{eqnarray}
&&
{\cal F}(a,T)=\frac{k_BT}{2\pi}\sum_{l=0}^{\infty}{\vphantom{\sum}}^{\prime}
\int_{0}^{\infty}\!\!\kb\,d\kb
\nonumber \\
&&
~~~\times\sum_{\alpha}
\ln\left[1-r_{\alpha}^{(1)}(i\xi_l,\kb)r_{\alpha}^{(2)}(i\xi_l,\kb)
e^{-2aq_l}\right],
\nonumber 
\end{eqnarray}
\begin{eqnarray}
&&
{P}(a,T)=-\frac{k_BT}{\pi}\sum_{l=0}^{\infty}{\vphantom{\sum}}^{\prime}
\int_{0}^{\infty}\!\!q_l\kb\,d\kb
\nonumber \\
&&
~~~\times\sum_{\alpha}
\left[\frac{e^{2aq_l}}{r_{\alpha}^{(1)}(i\xi_l,\kb)r_{\alpha}^{(2)}(i\xi_l,\kb)}
-1\right]^{-1}\!\!.
\label{eq1}
\end{eqnarray}

Here, $k_B$ is the Boltzmann constant, $\kb=(k_x^2+k_y^2)^{1/2}$ is the magnitude
of the wave vector projection on the plane of plates, $\xi_l=2\pi k_BTl/\hbar$
are the Matsubara frequencies, $q_l=(\skb+\xi_l^2/c^2)^{1/2}$, and the prime on
the summation sign divides by 2 the term of the first sums with $l=0$.
The reflection coefficients of electromagnetic waves with the transverse
magnetic ($\alpha={\rm TM}$) and transverse electric  ($\alpha={\rm TE}$)
polarizations on the first and second plates are $r_{\alpha}^{(1)}\xk$ and
$r_{\alpha}^{(2)}\xk$, respectively.

In the original version of the Lifshitz theory \cite{2,3}, it is assumed that the
plate materials possess only the temporal dispersion, i.e., their dielectric
permittivities, $\ve_n(\omega)$, and magnetic permeability, $\mu_n(\omega)$,
where $n=1,\,2$ for the first and second plates, depend on the frequency
$\omega$ (a dependence on $T$, e.g., for metals is also allowed).
In this case the reflection coefficients are given by the familiar Fresnel
formulas considered at $\omega=i\xi_l$
\begin{eqnarray}
&&
\rMn\xk=\frac{\ve_n(i\xi_l)q_l-k_n\xk}{\ve_n(i\xi_l)q_l+k_n\xk},
\nonumber \\
&&
\rEn\xk=\frac{\mu_n(i\xi_l)q_l-k_n\xk}{\mu_n(i\xi_l)q_l+k_n\xk},
\label{eq2}
\end{eqnarray}
\noindent
where
\begin{equation}
k_n\xk=\left[\skb+\ve_n(i\xi_l)\mu_n(i\xi_l)\frac{\xi_l^2}{c^2}
\right]^{1/2}\!\!.
\label{eq3}
\end{equation}

The reflection coefficients in Eq.~(\ref{eq1}) can also be expressed in terms of
the exact surface impedances $\zMn\xk$ and $\zEn\xk$
\begin{eqnarray}
&&
\rMn\xk=\frac{cq_l-\xi_l\zMn\xk}{cq_l+\xi_l\zMn\xk},
\nonumber \\
&&
\rEn\xk=\frac{cq_l\zEn\xk-\xi_l}{cq_l\zEn\xk+\xi_l},
\label{eq4}
\end{eqnarray}
\noindent
where for materials possessing only the temporal dispersion the impedances are
connected with the dielectric permittivities and magnetic permeabilities as
\cite{67,73}
\begin{eqnarray}
&&
\zMn\xk=\frac{ck_n\xk}{\xi_l\ve_n(i\xi_l)},
\nonumber \\
&&
\zEn\xk=\frac{\xi_l\mu_n(i\xi_l)}{ck_n\xk}.
\label{eq5}
\end{eqnarray}
\noindent
It is evident that the substitution of Eq.~(\ref{eq5}) in Eq.~(\ref{eq4}) returns
us back to the reflection coefficients (\ref{eq2}).

According to generalizations of the Lifshitz theory in the framework of the
scattering approach \cite{10,11,12}, Eq.~(\ref{eq1}) with appropriately defined
reflection coefficients remains valid for any planar structures.
If materials of the plates, besides temporal, possess the spatial dispersion,
derivation of the exact expressions for the reflection coefficients runs into
problems. The point is that the response of spatially dispersive material
filling in the entire 3-dimensional space to electric fields parallel and
perpendicular to the wave vector $\vk=(k_x,k_y,k_z)$ is described by the
longitudinal, $\Le(\omega,\vk)$, and transverse, $\Te(\omega,\vk)$, dielectric
permittivities \cite{65,74}.

In the strict sense, these permittivities can be introduced only under a condition
of translational invariance which is violated by the presence of Casimir plates
separated by a vacuum gap \cite{75,76,77}.
The standard Lifshitz theory deals with plate materials possessing only the
temporal dispersion. Therefore the dielectric permittivities depend only on
$\omega$ and the violation of translational invarianse makes no problem.
For the plate materials with spatial dispersion, this violation, however,
 makes impossible an employment
of the standard continuity boundary conditions and derivation of the Fresnel
reflection coefficients (\ref{eq2}).

This difficulty can be circumvented as follows.
 Since the Casimir force is determined by the dielectric properties
in the bulk and appropriate boundary conditions,
it is possible to preserve the translational invariance
in fictitious homogeneous medium
by assuming the specular reflection of charge carriers (electrons)
on the boundary surfaces of Casimir plates.
In doing so an electron reflected on an interface between the plate and the vacuum gap
is indistinguishable from an electron coming freely on the source side of a fictitious
medium.
Then one can introduce the longitudinal and transverse dielectric permittivities
$\ve^{\rm L}$ and $\ve^{\rm Tr}$ which depend on $\omega$ and all components
of the wave vector {\boldmath$k$} because the fictitious medium is translationally
invariant in all directions, and not only in the plane of Casimir plates \cite{61,62}.
Physically there is no sufficient reason to exclude one of the wave vector components.
This would be justified for graphene and other two-dimensional materials but not
for a bulky matter possessing the spatial dispersion.
Note that the response functions depending on all components of {\boldmath$k$}
are fully consistent with the Lifshitz theory where the reflection coefficients
depend only on $\kb$.
As a result, these
coefficients preserve the form of Eq.~(\ref{eq4}) whereas the surface impedances
are obtained from the nonlocal bulk permittivities depending  on $\omega$ and
 {\boldmath$k$}
by the integration with respect to $k_z$ as:
\begin{eqnarray}
&&\hspace*{-5mm}
\zMn\xk=\frac{c\xi_l\mu_n(i\xi_l)}{\pi}\int_{-\infty}^{\infty}
\frac{dk_z}{{\vk}^2}\biggl[
\frac{\skb}{\Lne(i\xi_l,\vk)\mu_n(i\xi_l)\xi_l^2}
\nonumber \\
&&~~~~
+\frac{k_z^2}{\Tne(i\xi_l,\vk)\mu_n(i\xi_l)\xi_l^2+c^2{\vk}^2}
\biggl],
\nonumber \\[-1mm]
&&~~~~
\label{eq6} \\[-1mm]
&&\hspace*{-5mm}
\zEn\xk=\frac{c\xi_l\mu_n(i\xi_l)}{\pi}\int_{-\infty}^{\infty}
\frac{dk_z}{\Tne(i\xi_l,\vk)\mu_n(i\xi_l)\xi_l^2+c^2{\vk}^2}.
\nonumber
\end{eqnarray}
\noindent
For nonmagnetic metals Eq.~(\ref{eq6}) was derived in Refs.~\cite{61,62}
(see also Ref.~\cite{78a} for a review) and generalized for the case
of magnetic ones in Ref.~\cite{73}.
There is also a generalization of Eq.~(\ref{eq6}) for the case of diffuse reflection \cite{66a,78b}.
Note, however, that for sufficiently smooth boundary surfaces used in experiments on
measuring the Casimir force the approximation of specular reflection is well applicable.

As mentioned in Sec.~I, the surface impedances (\ref{eq6})  with $\mu_n(i\xi_l)=1$
and the dielectric permittivities $\Le(\omega,\mbox{\boldmath$k$})$ and
$\Te(\omega,\mbox{\boldmath$k$})$ describing the anomalous skin
effect \cite{62} were used to calculate the Casimir force between Au surfaces in
Refs.~\cite{68,69}. It was found, however, that corrections to the force due to
spatial nonlocality in the region of the anomalous skin effect are too small and
incapable to  explain a disagreement between the measurement data and theoretical
predictions.

\section{Spatially nonlocal dielectric functions providing an alternative
response to the off-the-mass-shell fields}
\def\Xint#1{\mathchoice
   {\XXint\displaystyle\textstyle{#1}}%
   {\XXint\textstyle\scriptstyle{#1}}%
   {\XXint\scriptstyle\scriptscriptstyle{#1}}%
   {\XXint\scriptscriptstyle\scriptscriptstyle{#1}}%
   \!\int}
\def\XXint#1#2#3{{\setbox0=\hbox{$#1{#2#3}{\int}$}
     \vcenter{\hbox{$#2#3$}}\kern-.5\wd0}}
\def\ddashint{\Xint=}
\def\dashint{\Xint-}

The phenomenological nonlocal dielectric functions depending on $k_{\bot}$, which bring
the Lifshitz theory in agreement with experiments on measuring the Casimir force between
two similar plates and with thermodynamics do not disregarding the dissipation of
conduction electrons, were introduced in Refs.~\cite{71,72,73}.
Here, we propose the  nonlocal dielectric permittivities
depending on all components of the wave vector
\begin{eqnarray}
&&
\Tne(\omega,k)=\ve_c^{(n)}(\omega)-\frac{\omega_{p,n}^2}{\omega(\omega+i\gamma_n)}
\left(1+i\frac{\vTn k}{\omega}\right),
\nonumber \\[-1.5mm]
&&
\label{eq7}\\[-0.5mm]
&&
\Lne(\omega,k)=\ve_c^{(n)}(\omega)-\frac{\omega_{p,n}^2}{\omega(\omega+i\gamma_n)}
\left(1+i\frac{\vLn k}{\omega}\right)^{-1}\!\!\!\!\!,
\nonumber
\end{eqnarray}
\noindent
where $\ve_c^{(n)}(\omega)$ is the contribution determined by the core electrons,
$\omega_{p,n}$ is the plasma frequency, $\gamma_n$ is the relaxation parameter,
$k=|\vk|$ is the magnitude of the wave vector, and $\vTn$, $\vLn$ are constants
of the order of Fermi velocity $v_{\,{\rm F},n}$ (as before, $n=1,\,2$ for
materials of the first and second plates).

The response functions  introduced
in Refs.~\cite{71,72,73} are obtained from Eq.~(\ref{eq7}) if
 the magnitude of the wave vector
$\vk$ is replaced with the magnitude of its projection on the plane of the Casimir plates $\kb$.
 It was shown \cite{45,71} that the Lifshitz theory using the resulting
dielectric functions  is in  good agreement with the measurement data
of experiments \cite{27,36} measuring the Casimir interaction between two similar Au test
bodies if $v_{1,2}^{\rm Tr}=v_{1,2}^{\rm L}=7v_{\,\rm F,Au}$.
With the same numerical coefficient, $v_{1,2}^{\rm Tr}=v_{1,2}^{\rm L}=7v_{\,\rm F,Ni}$,
the Lifshitz theory using these permittivities was found in agreement \cite{73}
with measurements of the Casimir interaction between two similar magnetic (Ni) test
bodies \cite{31,32}. However, as discussed in previous section, the nonlocal permittivities
depending only on $\kb$ are not fully consistent with the used formalism of surface
impedances in the approximation of specular reflection and represent only some kind
of a simplified particular case.

We return to the permittivities (\ref{eq7}) which depend on  all components of the
wave vector.
For the electromagnetic waves on the mass shell (the propagating waves) they
 describe approximately the same response as the
standard Drude model supplemented by the oscillator terms
$\ve_c^{(n)}(\omega)$ taking into account
the interband transitions \cite{78}.  This is because the additions to unity
depending on $k$ in the parantheses in Eq.~(\ref{eq7}) under a condition $ck\leqslant\omega$
become negligibly small
\begin{equation}
\frac{v_n^{\rm Tr,L}k}{\omega}\sim \frac{v_{{\rm F},n}}{c}\,\frac{ck}{\omega}
\leqslant\frac{v_{{\rm F},n}}{c}\ll 1.
\label{eq8}
\end{equation}
\noindent
In doing so, the response functions (\ref{eq7}) take into account dissipation
of conduction electrons by means of the relaxation parameter $\gamma_n$
as does the Drude model.
If, however, $ck>\omega$, as it holds
for the evanescent waves  which are off the mass shell in free space, the permittivities
(\ref{eq7}) can lead to an electromagnetic response differing from that of the Drude
model.

Below we demonstrate that the response functions
(\ref{eq7}), which depend on all the three wave vector components, provide a complete
agreement with all measurements of the Casimir interaction including that one between
dissimilar (Au and Ni) plates \cite{30}. The case of dissimilar metals is of special interest
because in the range of experimental separations the Lifshitz theory using the Drude and
plasma dielectric permittivities at low frequencies leads to almost coinciding results which
are in agreement with the measurement data in the limits of the experimental errors
\cite{30}. Calculations show that the theoretical predictions using nonlocal dielectric
functions of Refs.~\cite{71,72,73} depending only on $\kb$ are also in agreement with
experiment if $v_{\rm Au}^{\rm Tr,L}=7v_{\,\rm F,Au}$ and
 $v_{\rm Ni}^{\rm Tr,L}=7v_{\,\rm F,Ni}$.
An advantage of the nonlocal response functions (\ref{eq7}) depending on all wave
vector components is, however, that they are not only fully consistent with the formalism
of surface impedances in the approximation of specular reflection, but lead to by a factor
of 4.7 smaller   $v_{\rm Au}^{\rm Tr,L}$ and $v_{\rm Ni}^{\rm Tr,L}$
(see Secs.~IV--VI). This significantly decreases any possible differences between theoretical
predictions obtained using the standard Drude model and its nonlocal modification in
optical experiments exploiting the propagating electromagnetic waves.

We begin with calculation of the reflection coefficients (\ref{eq4}) at zero
Matsubara frequency $\xi_0=0$ which plays the major role in the problems of
Casimir physics discussed in Sec.~I. Substituting $\omega=i\xi_l$ in Eq.~(\ref{eq7}),
one obtains the expressions for nonlocal permittivities (\ref{eq7}) at the pure
imaginary Matsubara frequencies
\begin{eqnarray}
&&
\Tne(i\xi_l,k)=\ve_c^{(n)}(i\xi_l)+\frac{\omega_{p,n}^2}{\xi_l(\xi_l+\gamma_n)}
\left(1+\frac{\vTn k}{\xi_l}\right),
\nonumber \\[-1.5mm]
&&
\label{eq9}\\[-0.5mm]
&&
\Lne(i\xi_l,k)=\ve_c^{(n)}(i\xi_l)+\frac{\omega_{p,n}^2}{\xi_l(\xi_l+\gamma_n)}
\left(1+\frac{\vLn k}{\xi_l}\right)^{-1}\!\!\!\!\!.
\nonumber
\end{eqnarray}

Substituting Eq.~(\ref{eq9}) to the first expression in Eq.~(\ref{eq6}) and considering vanishing $\xi_0$, we find the asymptotic behavior of $\zMn$ in this case
\begin{equation}
\zMn(i\xi_0,\kb)=\frac{2c\skb}{\pi\xi_0}\int_0^{\infty}\!\!\!
\frac{dk_z}{\sqrt{\skb+k_z^2}(\sqrt{\skb+k_z^2}+b_n)},
\label{eq10}
\end{equation}
\noindent
where $b_n\equiv\omega_{p,n}^2/(\gamma_n\vLn)$. Performing here the change
of integration variable from $k_z$ to $k=(\skb+k_z^2)^{1/2}$, we rewrite
Eq.~(\ref{eq10}) in the form
\begin{equation}
\zMn(i\xi_0,\kb)=\frac{2c\skb}{\pi\xi_0}\int_{\kb}^{\infty}\!\!\!
\frac{dk}{\sqrt{k^2-\skb}(k+b_n)}.
\label{eq11}
\end{equation}

This integral can be calculated using 1.2.45.3 and 1.2.45.5 in Ref.~\cite{79} with
the result
\begin{equation}
\zMn(i\xi_0,\kb)=\frac{2c\skb}{\pi\xi_0}
\left\{
\begin{array}{l}
\frac{1}{\sqrt{b_n^2-\skb}}\ln{\frac{b_n+\sqrt{b_n^2-\skb}}{\kb}},{\ }
\mbox{\scriptsize$\kb<b_n$}, \\
\frac{1}{\sqrt{\skb-b_n^2}}\arccos{\frac{b_n}{\kb}},{\ }
\mbox{\scriptsize$\kb>b_n$}.
\end{array}
\right.
\label{eq12}
\end{equation}

Substituting this result in the first expression of Eq.~(\ref{eq4}) and putting
$\xi_0=0$, we finally obtain
\begin{equation}
\rMn(0,\kb)=
\frac{\pi\sqrt{b_n^2-\skb}-2\kb\ln\frac{b_n+\sqrt{b_n^2-\skb}}{\kb}}{\pi
\sqrt{b_n^2-\skb}+2\kb\ln\frac{b_n+\sqrt{b_n^2-\skb}}{\kb}}
\label{eq13}
\end{equation}
\noindent
for $\kb<b_n$ and
\begin{equation}
\rMn(0,\kb)=
\frac{\pi\sqrt{\skb-b_n^2}-2\kb\arccos\frac{b_n}{\kb}}{\pi
\sqrt{\skb-b_n^2}+2\kb\arccos\frac{b_n}{\kb}}
\label{eq14}
\end{equation}
\noindent
for $\kb>b_n$. For $\kb=b_n$ both Eqs.~({\ref{eq13}) and (\ref{eq14}) lead
to the result
\begin{equation}
\rMn(0,\kb)=
\frac{\pi-2}{\pi+2}.
\label{eq15}
\end{equation}

Now we consider the value of the TE reflection coefficient at zero Matsubara
frequency. For this purpose, we substitute the first expression in Eq.~(\ref{eq9}) to
the second formula in Eq.~(\ref{eq6})
and find the following asymptotic expression in the
case of vanishing $\xi_0$:
\begin{equation}
\zEn(i\xi_0,\kb)=\frac{2\xi_0\mu_n(i\xi_0)}{\pi c}\int_0^{\infty}\!\!\!
\frac{dk_z}{B_n\sqrt{\skb+k_z^2}+\skb+k_z^2},
\label{eq16}
\end{equation}
\noindent
where $B_n\equiv\mu_n(i\xi_0)\omega_{p,n}^2\vTn/(\gamma_n c^2)$.

The integral in Eq.~(\ref{eq16}) has the same form as in Eq.~(\ref{eq10}).
Calculating it in the same way as above, one obtains
\begin{equation}
\zEn(i\xi_0,\kb)=\frac{2\xi_0\mu_n(i\xi_0)}{\pi c}
\left\{
\begin{array}{l}
\frac{1}{\sqrt{B_n^2-\skb}}\ln{\frac{B_n+\sqrt{B_n^2-\skb}}{\kb}},{\ }
\mbox{\scriptsize$\kb<B_n$}, \\
\frac{1}{\sqrt{\skb-B_n^2}}\arccos{\frac{B_n}{\kb}},{\ }
\mbox{\scriptsize$\kb>B_n$}.
\end{array}
\right.
\label{eq17}
\end{equation}

Substituting this equation in the second expression of Eq.~(\ref{eq4}) and putting
$\xi_0=0$, we find the following results:
\begin{equation}
\rEn(0,\kb)=
\frac{2\mu_n(0)\kb\ln\frac{B_n+\sqrt{B_n^2-\skb}}{\kb}-
\pi\sqrt{B_n^2-\skb}}{2\mu_n(0)\kb\ln\frac{B_n+\sqrt{B_n^2-\skb}}{\kb}+
\pi\sqrt{B_n^2-\skb}}
\label{eq18}
\end{equation}
\noindent
for $\kb<B_n$ and
\begin{equation}
\rEn(0,\kb)=
\frac{2\mu_n(0)\kb\arccos\frac{B_n}{\kb}-\pi\sqrt{\skb-B_n^2}}{2
\mu_n(0)\kb\arccos\frac{B_n}{\kb}+\pi\sqrt{\skb-B_n^2}}
\label{eq19}
\end{equation}
\noindent
for $\kb>B_n$. For $\kb=B_n$, Eqs.~(\ref{eq18}) and (\ref{eq19}) lead
to
\begin{equation}
\rEn(0,\kb)=
\frac{2\mu_n(0)-\pi}{2\mu_n(0)+\pi}.
\label{eq20}
\end{equation}

{}From Eqs.~(\ref{eq13}), (\ref{eq14}) and (\ref{eq18}), (\ref{eq19})
one can see that at zero Matsubara frequency the magnetic properties make
an impact only on the TE reflection coefficient in the Lifshitz formula
(\ref{eq1}).

The values of impedances and reflection coefficients at all Matsubara frequencies
with $l\geqslant 1$ are more complicated. We again substitute Eq.~(\ref{eq9}) in
the first expression in Eq.~(\ref{eq6}) and introduce in the obtained integrals
the following dimensionless integration variable and projection magnitude
of the wave vector on the plane of plates:
\begin{equation}
x=\frac{k_z c}{\xi_l}, \qquad
p_l=\frac{\kb c}{\xi_l}.
\label{eq21}
\end{equation}
\noindent
Then the impedance $Z_{\rm TM}$ takes the form
\begin{widetext}
\begin{eqnarray}
&&
\zMn\xk=\frac{2\skb c^2}{\pi\xi_l^2}\int_0^{\infty}\!\!\!\!
\frac{(c+\vLn\sqrt{p_l^2+x^2})dx}{(p_l^2+x^2)[cA_l^{(n)}+
\vLn\sqrt{p_l^2+x^2}]}
\label{eq22} \\
&&
+
\frac{2\mu_n(i\xi_l)}{\pi}\!\!\int_0^{\infty}\!\!\!\!
\frac{x^2\,dx}{(p_l^2+x^2)[\mu_n(i\xi_l)(A_l^{(n)}+
D_l^{(n)}\sqrt{p_l^2+x^2})+p_l^2+x^2]},
\nonumber
\end{eqnarray}
\end{widetext}
\noindent
where
\begin{equation}
A_l^{(n)}\equiv\ve_c^{(n)}(i\xi_l)+\frac{\omega_{p,n}^2}{\xi_l(\xi_l+\gamma_n)},
\quad
D_l^{(n)}\equiv\frac{\omega_{p,n}^2\vTn}{\xi_l(\xi_l+\gamma_n)c}.
\label{eq23}
\end{equation}

In a similar way, substituting the first expression in Eq.~(\ref{eq9}) in the
second formula of Eq.~(\ref{eq6}) and using Eq.~(\ref{eq21}), one obtains
\begin{widetext}
\begin{equation}
\zEn\xk=\frac{2\mu_n(i\xi_l)}{\pi}\int_0^{\infty}\!\!\!\!
\frac{dx}{\mu_n(i\xi_l)(A_l^{(n)}+
D_l^{(n)}\sqrt{p_l^2+x^2})+p_l^2+x^2}.
\label{eq24}
\end{equation}
\end{widetext}
\noindent
Equations (\ref{eq22}) and (\ref{eq24}) are convenient for numerical computations
performed in the next sections.

In the end of this section, we note that the spatially nonlocal dielectric
permittivities (\ref{eq7}) introduced above satisfy the Kramers-Kronig relations as
it should be in accordance with the condition of causality. This can be proven
similar to Ref.~\cite{71} if to take into account that the dielectric permittivity
of core electrons in Eq.~(\ref{eq7}) takes the form \cite{41,78}
\begin{equation}
\ve_c^{(n)}(\omega)=1+\sum_{j=1}^{K_n}
\frac{g_{n,j}}{\omega_{n,j}^2-\omega^2-i\gamma_{n,j}\omega},
\label{eq25}
\end{equation}
\noindent
where $\omega_{n,j}\neq 0$ are the oscillator frequencies, $g_{n,j}$ are
the oscillator strengths, $\gamma_{n,j}$ are the relaxation parameters and $K_n$
are the numbers of oscillators for the first ($n=1$) and second ($n=2$) Casimir
plates, respectively.  The permittivity (\ref{eq25}) satisfies the standard
Kramers-Kronig relations \cite{65,74}.

In this case the derivation presented in Ref.~\cite{71} is repeated with the only
replacement of $\kb$ with $k=(\skb+k_z^2)^{1/2}$. As a result, for $\Tne$ one
obtains the following Kramers-Kronig relations
\begin{eqnarray}
&&
{\rm Re\,}\Tne(\omega,k)=1+\frac{1}{\pi}\dashint_{-\infty}^{\infty}\!\!\!dx
\frac{{\rm Im\,}\Tne(x,k)}{x-\omega}
-\frac{\omega_{p,n}^2}{\omega^2}\,\frac{\vTn k}{\gamma_n},
\nonumber\\
&&
{\rm Im\,}\Tne(\omega,k)=-\frac{1}{\pi}\dashint_{-\infty}^{\infty}\!
\frac{dx}{x-\omega}\left[{\rm Re\,}\Tne(x,k)+
\frac{\omega_{p,n}^2}{x^2}\,\frac{\vTn k}{\gamma_n}\right]
\nonumber \\
&&~~~~~~~~~~~~~~~~~~
+\frac{4\pi{\rm Re\,}\sigma_{n,0}^{\rm Tr}(k)}{\omega},
\label{eq26}
\end{eqnarray}
\noindent
where the integrals are understood as the principal values and the real part
of the static transverse conductivity is given by \cite{71}
\begin{equation}
{\rm Re\,}\sigma_{n,0}^{\rm Tr}(k)=
\frac{\omega_{p,n}^2(\gamma_n-\vTn k)}{4\pi\gamma_n^2}.
\label{eq27}
\end{equation}

It should be mentioned that the last term on the right-hand side of the first
equality in Eq.~(\ref{eq26}) originates from the second-order pole of the
dielectric permittivity $\Tne(\omega)$ at $\omega=0$. The last term on the
right-hand side of the second  equality in Eq.~(\ref{eq26}) is caused by the
first-order pole so that in the local limit, $k\to 0$, Eq.~(\ref{eq27})
represents the static conductivity of the standard Drude model \cite{74}.

The dielectric permittivity $\Lne(\omega)$ defined in Eq.~(\ref{eq7}) has no poles at
$\omega=0$. For this reason, the Kramers-Kronig relations for this permittivity
take the same simplest form as for the permittivity of core electrons \cite{65,74}
\begin{eqnarray}
&&
{\rm Re\,}\Lne(\omega,k)=1+\frac{1}{\pi}\dashint_{-\infty}^{\infty}\!\!\!dx
\frac{{\rm Im\,}\Lne(x,k)}{x-\omega},
\nonumber \\
&&
{\rm Im\,}\Lne(\omega,k)=-\frac{1}{\pi}\dashint_{-\infty}^{\infty}\!\!\!
dx\frac{{\rm Re\,}\Lne(x,k)}{x-\omega}.
\label{eq28}
\end{eqnarray}

{}From Eqs.~(\ref{eq26}) and (\ref{eq28}), it is easy to find the expressions
for nonlocal dielectric permittivities along the imaginary frequency axis \cite{71,74}
\begin{eqnarray}
&&
\Tne(i\xi,k)=1+\frac{2}{\pi}\int_{0}^{\infty}\!\!\!dx
\frac{x{\rm Im\,}\Tne(x,k)}{x^2+\xi^2}
+\frac{\omega_{p,n}^2}{\xi^2}\,\frac{\vTn k}{\gamma_n},
\nonumber\\
&&
\Lne(i\xi,k)=1+\frac{2}{\pi}\int_{0}^{\infty}\!\!\!
dx\frac{x{\rm Im\,}\Lne(x,k)}{x^2+\xi^2},
\label{eq29}
\end{eqnarray}
\noindent
which are useful for computations by means of the Lifshitz formula (\ref{eq1}).

\section{Comparison with theory for measurements between nonmagnetic test
bodies by means of a micromechanical torsional oscillator}

In the series of dynamic experiments performed in high vacuum
in the configuration of an
Au-coated sphere of radius $R$ and an Au-coated plate by means of a micromechanical
torsional oscillator at room temperature \cite{25,26,27,28},
the measurement data for the gradient of the
Casimir force $F_{sp}^{\prime}(a,T)$ was represented in terms of the effective
Casimir pressure between two Au plates
\begin{equation}
P^{\rm expt}(a,T)=-\frac{1}{2\pi R}F_{sp}^{\prime}(a,T).
\label{eq30}
\end{equation}
\noindent
This was done by using the proximity force approximation \cite{40,41} which leads
to a relative error of less than $a/R$ (all measurements were performed at the
sphere-plate separations $a<750~$nm with a sphere radius $R=151.2~\mu$m \cite{27}).

It was found \cite{25,26,27,28} that the theoretical predictions of the Lifshitz
theory computed by Eq.~(\ref{eq1}) with taken into account surface roughness
\cite{40,41,80,81,82} are excluded by the measurement data if the dielectric
response of Au is described by the Drude model supplemented by the permittivity
of core electrons. This response function is obtained from Eq.~(\ref{eq9}) with
$k=0$
\begin{equation}
\ve_n^{D}(i\xi_l)=\Tne(i\xi_l,0)=\Lne(i\xi_l,0)=
\ve_c^{(n)}(i\xi_l)+\frac{\omega_{p,n}^2}{\xi_l(\xi_l+\gamma_n)}.
\label{eq31}
\end{equation}
\noindent
In doing so the values of $\ve_c^{(n)}(i\xi_l)$ were found from the optical data
for the complex index of refraction of Au \cite{83} using the Kramers-Kronig
relations. If, however, the response function (\ref{eq31}) with $\gamma_n=0$
is used in computations (i.e., the plasma-like model which disregards the
relaxation properties of conduction electrons),
 the theoretical predictions turn out to be in good agreement
with the measurement data \cite{25,26,27,28}.

Here, we compare the measurement data of the most precise experiment of these
series \cite{27} and theoretical predictions of the Lifshitz theory using the
nonlocal dielectric permittivities (\ref{eq9}). Numerical computations of the
Casimir pressure were performed by Eqs.~(\ref{eq1}), (\ref{eq4}), (\ref{eq6}),
and (\ref{eq9}). In so doing the reflection coefficients at zero Matsubara
frequency were calculated by Eqs.~(\ref{eq13}), (\ref{eq14}) and (\ref{eq19}),
(\ref{eq20}). The values of impedances at all Matsubara frequencies with
$l\geqslant 1$ were computed by Eqs.~(\ref{eq22}) and (\ref{eq24}).
The surface roughness was accounted for in the same way as in Ref.~\cite{27},
i.e., with the help of an additive approach. Its contribution to the effective
Casimir pressure turns out to be negligibly small.

The following values of all parameters have been used in computations. Taking into
account that both test bodies were made of Au, one should put
$\hbar\omega_{p,1}=\hbar\omega_{p,2}=\hbar\omega_{p,{\rm Au}}=8.9~$eV
and $\hbar\gamma_1=\hbar\gamma_2=\hbar\gamma_{\rm Au}=35.7~$meV \cite{27,40,41}.
Since Au is a nonmagnetic metal, we have $\mu_1(i\xi_l)=\mu_2(i\xi_l)=1$.
The values of $\ve_c^{(1)}(i\xi_l)=\ve_c^{(2)}(i\xi_l)=\ve_c^{(Au)}(i\xi_l)$
were calculated in Refs.~\cite{25,26,27,28} using the tabulated optical data for
the complex index of refraction of Au \cite{83} and used here.
For the constants of the order of Fermi velocity in Eqs.~(\ref{eq7}) and (\ref{eq9})
we have used the equal values $\vTn=\vLn=v_n$. In so doing, in this experiment
and in all other experiments considered below, the best agreement between
experiment and theory is reached for $v_n=3v_{{\rm F},n}/2$ with the value of
$v_{{\rm F},n}$ found under an assumption of the spherical Fermi surface
\begin{equation}
\frac{m_ev_{{\rm F},n}^2}{2}=\hbar\omega_{p,n},
\label{eq32}
\end{equation}
\noindent
where $m_e$ is the electron mass. Thus, for Au in this experiment from
Eq.~(\ref{eq32}) one obtains
$v_{{\rm F},1}=v_{{\rm F},2}=v_{{\rm F, Au}}=1.77\times 10^6~$m/s.
The numerical factor between $v_n$ and $v_{{\rm F},n}$, which is equal to 3/2
for the dielectric permittivities (\ref{eq7}) and (\ref{eq9}), is in fact the single
fitting parameter of our phenomenological model. The chosen value of this parameter
leads to the best agreement between experiment and theory. We recall that for the
nonlocal permittivities of Refs.~\cite{71,72,73} depending only on $\kb$ this
parameter is equal to 7 resulting in larger (but as yet sufficiently small) deviations
from the standard Drude model in optical experiments.

\begin{figure*}[!t]
\vspace*{2cm}
\centerline{
\includegraphics[width=6.50in]{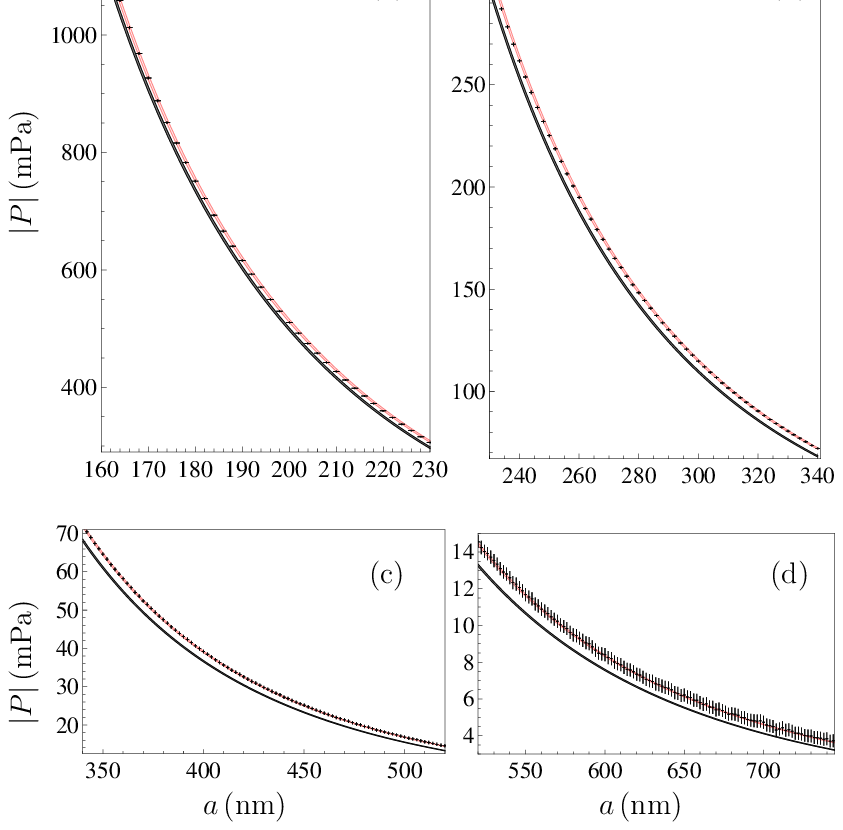}}
\vspace*{-11.cm}
\caption{\label{fg1}
The magnitudes of the effective Casimir pressure between two Au-coated
parallel plates computed using the spatially nonlocal response functions
and the Drude model are shown as functions of separation by the upper (red) and
lower (black) bands, respectively. The experimental data \cite{27}
are shown as crosses whose
arms indicate the total errors determined at the 95\% confidence level.
}
\end{figure*}
The computational results for the magnitude of the effective Casimir pressure
are shown by the upper (red) theoretical bands in Figs.~\ref{fg1}(a)--\ref{fg1}(d)
over four separation regions from 162.03 to 745.98~nm. The experimental data
are shown as crosses. The horizontal and vertical arms of these crosses indicate
the total measurement errors found at the 95\% confidence level as
a combination of systematic and random errors \cite{27}. The lower (black) theoretical
bands in Fig.~\ref{fg1} indicate the theoretical predictions of the Lifshitz theory
obtained by Eqs.~(\ref{eq1}) and (\ref{eq2}) using the standard local Drude response
function (\ref{eq31}). The width of each theoretical band is determined by the
errors in all the above theoretical parameters used in computations.

As is seen in Fig.~\ref{fg1}, the measurement data are in a very good agreement
with theoretical predictions of the Lifshitz theory made using the spatially
nonlocal response functions (\ref{eq7}), (\ref{eq9}) which take into account
the dissipation properties of conduction electrons.  Almost the same theoretical
predictions in agreement with the measurement data were made by the Lifshitz
theory using the plasma response function given by Eq.~(\ref{eq31}) with
$\gamma_n=0$, i.e., with the relaxation properties of free electrons disregarded.
The respective theoretical bands cannot be distinguished from the upper (red) bands
shown in Fig.~\ref{fg1}. As to the lower (black) theoretical bands in Fig.~\ref{fg1}
computed using the Drude  response function (\ref{eq31}), they are excluded
by the measurement data over the entire range of separations.

We are coming now to the recently performed experiment on measuring the
differential Casimir force between an Au-coated sphere of $R=149.7~\mu$m radius
and top and bottom of the Au-coated rectangular trenches \cite{37}.
As most of precise measurements of the Casimir interaction, this one was made
at room temperature in high vacuum. Thanks to the differential character of
this measurement, it has been made possible to obtain the meaningful data up
to separation distances of a few micrometers using the same setup of a
micromechanical torsional oscillator. Due to the sufficiently deep trenches
used, the effectively measured Casimir force was that acting between a sphere
and a plate which served as the trench top.

In the framework of the proximity force approximation, the Casimir force acting
between a sphere and a plate is given by
\begin{equation}
F_{sp}(a,T)=2\pi R{\cal F}(a,T),
\label{eq33}
\end{equation}
\noindent
where the Casimir free energy in the configuration of two parallel plates
is presented in Eq.~(\ref{eq1}). In Ref.~\cite{37}, the force $F_{sp}$ was
computed both approximately using Eqs.~(\ref{eq1}), (\ref{eq2}), and (\ref{eq33})
and precisely on the basis of first principles of quantum electrodynamics at
nonzero temperature using the scattering approach \cite{84,85,86,87} and the
gradient expansion \cite{88,89,90,91}. It was shown \cite{37} that all differences
between the approximate and exact results are well below the measurement errors
within the separation region from 0.2 to $8~\mu$m, irrespective of whether the
Drude or plasma response function is used in computations.

The obtained theoretical results employing the plasma model given by Eq.~(\ref{eq31})
with $\gamma_n=0$ were found to be in a good agreement with the measurement data
over the entire range of separations. The results computed similarly, but using
the Drude model, were excluded by the data over the separation region
from 0.2 to $4.8~\mu$m. In so doing the background electric force due to patch
potentials was investigated with the help of Kelvin probe microscopy \cite{92}
and included in the total experimental error of the Casimir force determined
at the 95\% confidence level.

Here, we compute the Casimir force in the configuration of the experiment \cite{37}
by the first equality in Eq.~(\ref{eq1}) and Eq.~(\ref{eq33}) using the proposed
spatially nonlocal dielectric functions (\ref{eq9}). For the reflection coefficients
with $l=0$, Eqs.~(\ref{eq13}), (\ref{eq14}) and (\ref{eq19}), (\ref{eq20}) have been
used, and for $l\geqslant 1$ numerical computations were performed by
Eqs.~(\ref{eq22}) and (\ref{eq24}) as described above with the following values of
all parameters of Au \cite{37}: $\hbar\omega_{p,{\rm Au}}=9.0~$eV,
$\hbar\gamma_{\rm Au}=35.0~$meV, $v_{\rm F,Au}=1.78\times 10^6~$m/s.
The obtained results were multiplied by a correction factor accounting for an
inaccuracy of the proximity force approximation which was calculated in
Ref.~\cite{37}.

\begin{figure}[!b]
\vspace*{-2cm}
\centerline{
\includegraphics[width=4.50in]{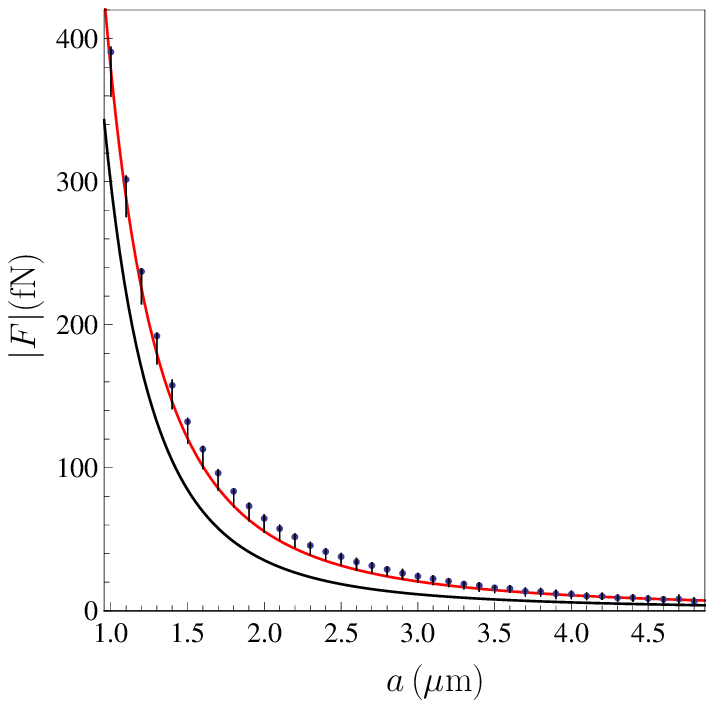}}
\vspace*{-8.cm}
\caption{\label{fg2}
The magnitudes of the Casimir force between an Au-coated sphere and an
Au-coated plate computed using the spatially nonlocal response functions
and the Drude model are shown as functions of separation by the upper (red) and
lower (black) bands, respectively. The experimental data \cite{37}
are shown as crosses whose
arms indicate the total errors determined at the 95\% confidence level.
}
\end{figure}
The obtained computational results in the range of separations from 1 to $4.9~\mu$m
are shown by the upper (red) band in Fig.~\ref{fg2}. In the same figure, the lower
(black) band
is computed by Eqs.~(\ref{eq1}), (\ref{eq2}) and (\ref{eq33}) using the Drude
response function (\ref{eq31}). The measurement data are indicated as crosses.
As is seen in Fig.~\ref{fg2}, the upper and lower vertical arms of the crosses
differ from one another. This is because the attractive electric force due to patch
potentials is included as part of the error in measuring the Casimir force.
According to Fig.~\ref{fg2}, the theoretical predictions of the Lifshitz theory
using the proposed nonlocal response functions (\ref{eq7}), (\ref{eq9}) are in
agreement with the measurement data.
Good agreement also holds over the ranges of separations from 0.2 to $1~\mu$m and
from 4.8 to $8~\mu$m which are not shown in Fig.~\ref{fg2}.
The predictions of the same theory using the
Drude model are excluded over the range of separations from 1 to $4.8~\mu$m
(in Ref.~\cite{37} it is shown that they are also excluded in the measurement range
from 0.2 to $1~\mu$m but here we are more interested in the region of large
separations exceeding $1~\mu$m).

The Casimir forces computed by Eqs.~(\ref{eq1}), (\ref{eq2}) and (\ref{eq33}) using
the plasma model given by  Eq.~(\ref{eq31}) with $\gamma_n=0$ are almost the same
as the ones computed above with the spatially nonlocal response functions.
Thus, at separations of 1, 3, 5, and $7~\mu$m the pairs of force magnitudes (in fN)
computed using the nonlocal response functions and the plasma model are
(372.98, 374.62), (21.44, 21.67), (7.24, 7.30), and (3.69, 3.71), i.e., only 0.44\%,
1.06\%, 0.8\%, and 0.54\% relative differences, well below the respective
experimental errors. Although these results are not experimentally distinguishable,
that ones obtained using the nonlocal response functions should be considered as
preferable as they are obtained with taken into account relaxation properties of
conduction electrons.

\section{Measurements between nonmagnetic test
bodies by means of an atomic force microscope}
Another experimental setup for measuring the Casimir interaction is an atomic force
microscope whose sharp tip is replaced with a sphere of sufficiently large radius
 \cite{93}. Here, we compare the measurement data of three most precise experiments
on measuring the gradient of the Casimir force between an Au-coated sphere and an
Au-coated plate obtained by means of a dynamic atomic force microscope \cite{29,36}
with theoretical predictions using the nonlocal dielectric functions (\ref{eq9}).
All measurements were performed in high vacuum at room temperature.
In interpretation of all these experiments the same parameters of Au, i.e., the
values of $\omega_{p,{\rm Au}}$, $\gamma_{\rm Au}$, and $\ve_c^{(\rm Au)}$,
have been used as already listed in Sec.~IV when describing measurements of the
differential Casimir force between a sphere and a plate with rectangular trenches.

We start with the experiment of Ref.~\cite{29} which employed the sphere of
$R=41.3~\mu$m radius. The theoretical force gradients were computed using the
Lifshitz theory and the proximity force approximation with taken into account
correction for its inaccuracy \cite{29}
\begin{equation}
F_{sp}^{\prime}(a,T)=-2\pi RP(a,T)\left[1+\theta(a,T)\frac{a}{R}\right],
\label{eq34}
\end{equation}
\noindent
where the Casimir pressure $P$ is given by the second expression in Eq.~(\ref{eq1})
with the Fresnel reflection coefficients (\ref{eq2}) and at separations below
$1~\mu$m the coefficient $\theta$ is negative and does not exceed unity (see
Refs.~\cite{88,89,90} and the more complete results for different dielectric functions
in Refs.~\cite{87,91}). The effect of surface roughness was taken into account perturbatively and shown to be negligibly small.

According to the results of Ref.~\cite{29}, the theoretical predictions obtained
using the Drude model (\ref{eq31}) are excluded by the measurement data within
the range of separations from 235 to 420~nm. The same data turned out to be
in a good agreement with theoretical predictions found by using the plasma model
which does not take into account the relaxation properties of free electrons, i.e.,
put $\gamma_{\rm Au}=0$. Thus, the results obtained earlier in
Refs.~\cite{25,26,27,28} by means a micromechanical torsional oscillator were
confirmed independently by using quite different laboratory setup.

\begin{figure}[!b]
\vspace*{-3.5cm}
\centerline{\hspace*{3cm}
\includegraphics[width=6.50in]{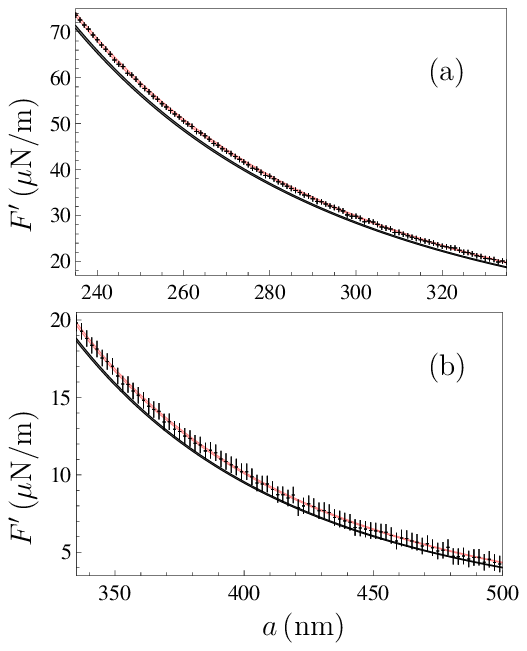}}
\vspace*{-11.cm}
\caption{\label{fg3}
The gradients of the Casimir force between an Au-coated sphere and an
Au-coated plate computed using the spatially nonlocal response functions
and the Drude model are shown as functions of separation by the upper (red)  and
lower (black) bands, respectively. The experimental data \cite{29}
are shown as crosses whose
arms indicate the total errors determined at the 67\% confidence level.
}
\end{figure}
We have computed the gradients of the Casimir force (\ref{eq34}) in the experimental
configuration of Ref.~\cite{29} using the spatially nonlocal response functions
(\ref{eq9}) and reflection coefficients (\ref{eq4}) expressed via the surface
impedances as described in Secs.~III and IV. The same parameters of Au, as in
Ref.~\cite{29}, have been used and
$v_n^{\,\rm Tr,L}=v_{\rm Au}^{\,\rm Tr,L}=3v_{\rm F,Au}/2$ as indicated above.
The computational results are shown by the upper (red) bands in Fig.~\ref{fg3} where
the experimental data are presented as crosses whose arms indicate the total
measurement errors determined at the 67\% confidence level. The lower (black) bands
in Fig.~\ref{fg3} reproduce the computational results of Ref.~\cite{29} obtained
using the Drude model (\ref{eq31}). In fact, our computational results using the
nonlocal response functions are almost coinciding with the results of
Ref.~\cite{29} using the dissipationless plasma model. In doing so, our results
are in a good agreement with the measurement data over the entire range of
separations which exclude the theoretical predictions using the Drude model over
the separation range  from 235 to 420~nm.

An upgraded setup employing the atomic force microscope with increased sensitivity of
the cantilever through a decrease of its spring constant was used in
Refs.~\cite{35,36} in more precise measurements of the Casimir force gradients
up to larger separation distances. The important property of an upgraded setup
was an employment of the two-step cleaning procedure of the vacuum chamber and
test body surfaces by means of UV light followed by Ar-ion bombardment.
The radius of the sphere used was $R=43.47~\mu$m.

The comparison between experiment and theory in Ref.~\cite{36} was made as
described above in this section using Eq.~(\ref{eq34}) and the dielectric response
(\ref{eq31}) with $\gamma_{\rm Au}\neq 0$ (the Drude model) and
$\gamma_{\rm Au}= 0$ (the plasma model). In the measurements with smaller oscillation
amplitude of the cantilever equal to 10~nm, the theoretical predictions using the
Drude model were excluded over the range of separations from 250 to 820~nm
whereas those using the plasma model were found to be in agreement with the
measurement data over the entire measurement range.

We have computed the gradient of the Casimir force in the experimental configuration
of Refs.~\cite{35,36} using the proposed dielectric functions (\ref{eq9}) with the
same parameters of Au, as in these references, and
$v_{\rm Au}^{\,\rm Tr,L}=3v_{\rm F,Au}/2$ as above. The computational results are
shown by the upper (red)  bands in Fig.~\ref{fg4} as a function of separation.
The measurement data with their errors determined at the 67\% confidence level
are shown as crosses. The theoretical predictions obtained using the Drude model
\cite{36} are presented by the lower (black) bands. As is seen in Fig.~\ref{fg4}, our
results, which take into account the dissipation of free electrons, are in
agreement with the data over the entire separation region from 250 to 950~nm.

\begin{figure}[!b]
\vspace*{-4.cm}
\centerline{\hspace*{3cm}
\includegraphics[width=6.50in]{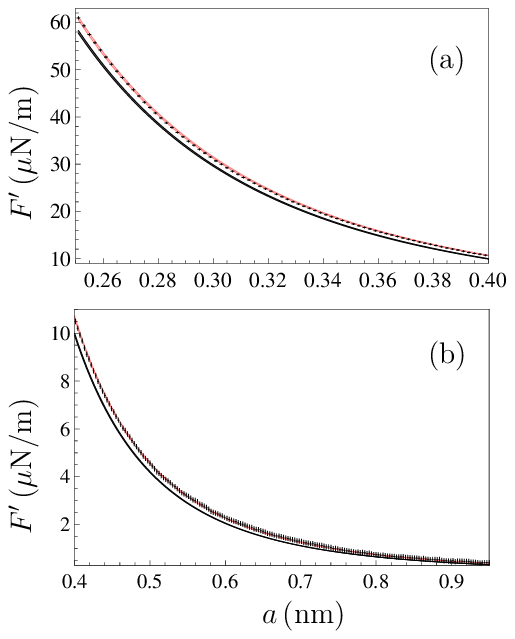}}
\vspace*{-11.cm}
\caption{\label{fg4}
The gradients of the Casimir force between an Au-coated sphere and an
Au-coated plate computed using the spatially nonlocal response functions
and the Drude model are shown as functions of separation by the upper (red) and
lower (black) bands, respectively. The experimental data \cite{36} obtained with
smaller oscillation amplitude of the cantilever are shown as crosses whose
arms indicate the total errors determined at the 67\% confidence level.
}
\end{figure}
Another set of measurements was performed in Ref.~\cite{36} with a larger
oscillation amplitude of cantilever equal to 20~nm. This made it possible
to get the meaningful measurement data at larger separation distances and exclude
theoretical predictions using the Drude model up to $1.1~\mu$m.

Our computational results for the gradient of the Casimir force
obtained with the nonlocal
dielectric functions (\ref{eq9}) are presented by the upper (red) bands in Fig.~\ref{fg5}
over the range of separations from 0.6 to $1.3~\mu$m of the data set in
Ref.~\cite{36} measured with a larger oscillation amplitude. They are in a
good agreement with the measurement data indicated as crosses. Our results
accounting for the dissipation properties of free electrons are almost
coinciding with those obtained in Ref.~\cite{36} using the dissipationless
plasma model, but deviate significantly from those obtained by means of the
Drude model. The latter are shown by the lower (black) bands.

One can conclude that the Lifshitz theory employing the proposed nonlocal
dielectric permittivity is in equally good agreement with the measurement data
of precise experiments performed by two different experimental groups by means
of a micromechanical torsional oscillator and an atomic force microscope using
test bodies made of a nonmagnetic metal.
\begin{figure}[!b]
\vspace*{-4cm}
\centerline{\hspace*{3cm}
\includegraphics[width=6.50in]{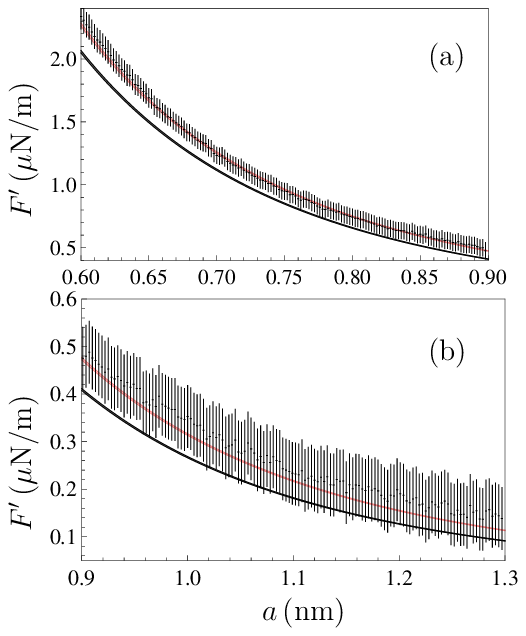}}
\vspace*{-11.cm}
\caption{\label{fg5}
The gradients of the Casimir force between an Au-coated sphere and an
Au-coated plate computed using the spatially nonlocal response functions
and the Drude model are shown as functions of separation by the upper (red)  and
lower (black) bands, respectively. The experimental data \cite{36} obtained with
larger oscillation amplitude of the cantilever are shown as crosses whose
arms indicate the total errors determined at the 67\% confidence level.
}
\end{figure}

\section{Theory-experiment comparison with magnetic test
bodies }

The action of magnetic properties of the plate materials on the Casimir force
has attracted considerable attention in the literature. Thus, although in
Refs.~\cite{2} and \cite{3} the Lifshitz formulas were derived for nonmagnetic
test bodies, they were rewritten with account of magnetic properties in
Ref.~\cite{94}. The Casimir force acting between an ideal metal plate and
an infinitely permeable one was found in  Ref.~\cite{95}.
Thereafter the Casimir force between one magnetic and one nonmagnetic plates,
as well as between two magnetic plates, was considered by many authors (see, e.g.,
Refs.~\cite{12,75,96,97,98}). This problem attracted special attention in
connection with the possibility of repulsive Casimir forces \cite{42}.

As was emphasized in Ref.~\cite{99}, the magnetic properties of boundary plates make
an impact on the Casimir free energy and pressure entirely through the
zero-frequency terms of the Lifshitz formulas (\ref{eq1}). This is caused by the
fact that the frequency-dependent magnetic permeability becomes equal to unity
at frequencies which are much smaller than the first Matsubara frequency at not
too low temperature \cite{100,101}.

We begin with an experiment of Ref.~\cite{30} where the gradient of the Casimir
force was measured between an Au-coated sphere of $R=64.1~\mu$m radius and a plate
coated with a magnetic metal Ni by means of an atomic force microscope.
Note that in measuring of the Casimir force using magnetic metals they are not
magnetized and do not give rise to a gradient of any additional force of
magnetic origin \cite{32}.
In Ref.~\cite{30} the theoretical force gradients were computed by Eq.~(\ref{eq34})
with the Fresnel reflection coefficients (\ref{eq2}) and spatially local dielectric
permittivities
(\ref{eq31}) with $\gamma_{\rm Au}\neq 0$ (the Drude model) and
$\gamma_{\rm Au}= 0$ (the plasma model) at room temperature.
The following values of parameters for Au ($n=1$) and Ni ($n=2$) have been used:
$\hbar\omega_{p,1}=\hbar\omega_{p,{\rm Au}}=9.0~$eV,
$\hbar\gamma_{1}=\hbar\gamma_{\rm Au}=35.0~$meV \cite{40,41,83} and
$\mu_1(i\xi_l)=\mu_{\rm Au}(i\xi_l)=1$;
$\hbar\omega_{p,2}=\hbar\omega_{p,{\rm Ni}}=4.89~$eV,
$\hbar\gamma_{2}=\hbar\gamma_{\rm Ni}=43.6~$meV \cite{83,102} and
$\mu_2(0)=\mu_{\rm Ni}(0)=110$,
$\mu_2(i\xi_l)=\mu_{\rm Ni}(i\xi_l)=1$ for $l\geqslant 1$.
The values of $\ve_c^{(1)}(i\xi_l)=\ve_c^{(\rm Au)}(i\xi_l)$ and
$\ve_c^{(2)}(i\xi_l)=\ve_c^{(\rm Ni)}(i\xi_l)$ were obtained from the optical data
of Au and Ni, respectively, as described in Refs.~\cite{29,30,32,40,41}.

According to the results of Ref.~\cite{30}, within the measurement range from 220
to 500~nm the theoretical predictions of the Lifshitz theory using the Drude and
the plasma models are almost coinciding and are in a good agreement with the
measurement data.
(Note that at larger separations of a few micrometers the gradients of the Casimir
force between a nonmagnetic-metal sphere and a magnetic-metal plate computed using
the Drude and plasma models are different \cite{99}.)
The predictions obtained by means of the plasma
model with magnetic properties of the Ni plate disregarded
[$\mu_{\rm Ni}(i\xi_l)=1$ for all $l\geqslant 0$] are excluded by the measurement
data over the range of separations from 220 to 420~nm. If the Drude model is used
in computations, the obtained results do not depend on whether the magnetic
properties of Ni plate are included or omitted.

Here, we compute the gradient of the Casimir force in the experimental configuration
of Ref.~\cite{30} using Eq.~(\ref{eq34}) and the spatially nonlocal dielectric
permittivities (\ref{eq9}). When computing the Casimir pressure, we have used the
impedance reflection coefficients (\ref{eq4}) for Au and Ni leading to
Eqs.~(\ref{eq13}), (\ref{eq14}) and (\ref{eq18}), (\ref{eq19}) for $l=0$ and
Eqs.~(\ref{eq22}), (\ref{eq24}) for $l\geqslant 1$.
The same parameters for Au and Ni, as indicated above have been used and the Fermi
velocity for Ni was found from Eq.~(\ref{eq32}) under an assumption of the spherical
Fermi surface, $v_{\rm F,Ni}=1.31\times 10^6~$m/s (for Au we used
$v_{\rm F,Au}=1.78\times 10^6~$m/s employed in Sec.~IV and V).
As in all previously considered experiments, the best agreement between the
measurement data and theoretical predictions is reached for
$v_{\rm Au}^{\rm Tr}=v_{\rm Au}^{\rm L}=3v_{\rm F,Au}/2$ and
$v_{\rm Ni}^{\rm Tr}=v_{\rm Ni}^{\rm L}=3v_{\rm F,Ni}/2$.

\begin{figure}[!b]
\vspace*{-4cm}
\centerline{\hspace*{3cm}
\includegraphics[width=6.50in]{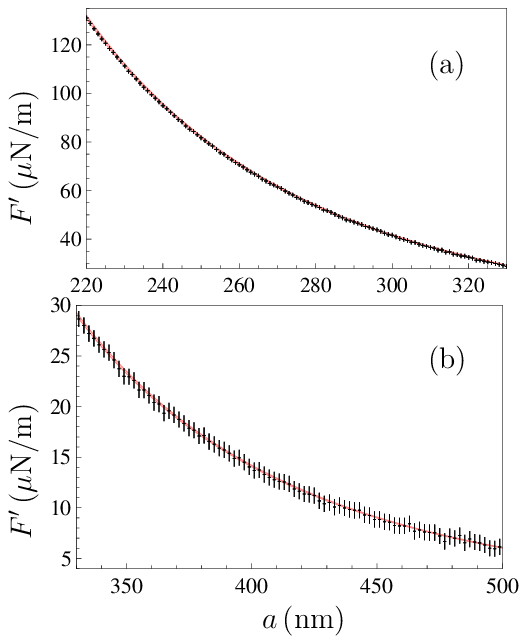}}
\vspace*{-11.cm}
\caption{\label{fg6}
The gradients of the Casimir force between an Au-coated sphere and a
Ni-coated plate computed using the spatially nonlocal response functions
 are shown by the solid (red) bands as a function of separation.
 The experimental data \cite{30}  are shown as crosses whose
arms indicate the total errors determined at the 67\% confidence level.
}
\end{figure}
Computational results obtained using the spatially nonlocal dielectric
permittivities (\ref{eq9}) are shown by the solid (red) bands in Fig.~\ref{fg6}
where the experimental data with their total errors determined at the 67\%
confidence level are presented as crosses. As is seen in Fig.~\ref{fg6},
the theoretical predictions of the Lifshitz theory employing the proposed
nonlocal dielectric functions are in a very good agreement with the measurement
data over the entire range of experimental separations.
For the configuration of Au-Ni test bodies in the range from 220 to 500~nm
almost the same theoretical predictions in equally good agreement with the
measurement data are obtained when the conduction electrons are described
by the spatially local Drude or plasma model.

Next, we consider the experiment of Refs.~\cite{31,32} on measuring the gradient
of the Casimir force between a sphere of $R=61.71~\mu$m radius and a plate both
coated with a magnetic metal Ni performed by means of an atomic force microscope.
In these references computations of the force gradient using the standard Lifshitz
theory were performed as in Sec.~V but with the parameters
$\omega_{p,1}=\omega_{p,2}=\omega_{p,{\rm Ni}}$,
$\gamma_{1}=\gamma_{2}=\gamma_{\rm Ni}$, and
$\ve_c^{(1)}(i\xi_l)=\ve_c^{(2)}(i\xi_l)=\ve_c^{(\rm Ni)}(i\xi_l)$
presented above. An important result found in Ref.~\cite{31} is that for two
magnetic metals the gradients of the Casimir force calculated using the Drude
model are larger than those found by means of the plasma model.  This is
different from the case of two Au plates (see Figs.~\ref{fg3}--\ref{fg5}).
It was shown \cite{31,32} that the theoretical predictions obtained with the
Drude model are excluded by the data over the separation region from 223 to
420~nm, whereas similar predictions made with the help of a plasma model are
in a very good agreement with the measurement results. Similar to the case
of test bodies coated with Au films, this result is puzzling because at low
frequencies the relaxation properties of conduction electrons are well
studied in many physical phenomena other than the Casimir effect.

\begin{figure}[!t]
\vspace*{-4cm}
\centerline{\hspace*{3cm}
\includegraphics[width=6.50in]{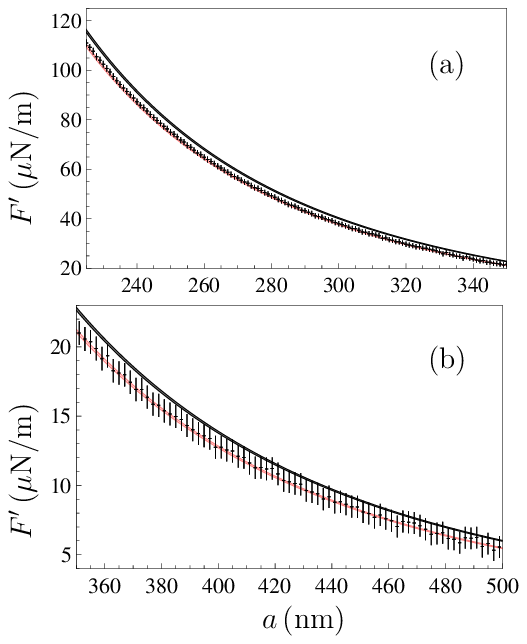}}
\vspace*{-11.cm}
\caption{\label{fg7}
The gradients of the Casimir force between a Ni-coated sphere and a
Ni-coated plate computed using the spatially nonlocal response functions
and the Drude model are shown as functions of separation by the
lower (red) and upper (black) bands, respectively. The experimental data \cite{31,32}
 are shown as crosses whose
arms indicate the total errors determined at the 67\% confidence level.
}
\end{figure}
Here, we computed the gradient of the Casimir force between the Ni-coated
surfaces of a sphere and a plate by Eq.~(\ref{eq34}).
The Casimir pressure in this equation was found from the second line in
Eq.~(\ref{eq1}), reflection coefficients (\ref{eq4}), and impedance functions
(\ref{eq6}), as described above, using the spatially nonlocal dielectric
permittivities (\ref{eq9}) with the same parameters of Ni as above.
The computational results are shown by the lower (red) bands in Fig.~\ref{fg7} as
a function of separation. They are in a very good agreement with the measurement
data indicated as crosses. As in all other experiments using an atomic force
microscope, the total experimental errors are determined at the 67\% confidence
level. The upper (black) bands in Fig.~\ref{fg7} reproduce the results of
Refs.~\cite{31,32} computed by the Lifshitz formula and the spatially local
Drude model (\ref{eq31}). It is seen that these results are excluded by the
measurement data over the separation region from 223 to 420~nm in accordance with
the conclusion made in Refs.~\cite{31,32}. The point is that at short separations
up to a few hundred nanometers the force gradients computed using the local plasma
model, which disregards the relaxation of free electrons, and the spatially
nonlocal dielectric functions (\ref{eq9}), which take relaxation into account,
are almost coinciding. In this situation, a failure of the Drude model may be
explained by an inadequate description of the dielectric response to
electromagnetic waves off the mass shell.

\section{Conclusions and discussion}

In this paper, we have proposed the phenomenological spatially nonlocal dielectric
functions which provide nearly the same response to electromagnetic waves on the
mass shell, as does the standard Drude model, but respond differently to the
off-the-mass-shell fields. Unlike the previously suggested response functions
of this kind \cite{71,72,73}, the permittivities presented here depend on all the
three components of the wave vector which is a more general case in the approximation
of  specular reflection used.

As discussed in Sec.~I, the problem of disagreement between theoretical predictions
of the fundamental Lifshitz theory using the well-tested Drude model with the
experimental data for metallic test bodies remains unresolved for almost 20 years.
Many attempts of its resolution have been undertaken (including a consideration of
the frequency-dependent relaxation parameter \cite{103}, i.e., the so-called
Gurzhi model), but the problem is as yet unresolved. Similar problem arises for
dielectric materials \cite{45,104}. All this makes it warranted to consider some
phenomenological approaches suggested by analogy with graphene which, due to its
simplicity in comparison with metallic materials, allows the fundamental calculation
of its spatially nonlocal dielectric response based on the first principles of
quantum electrodynamics at nonzero temperature.

Using this line of reasoning, the surface impedances and reflection coefficients
determined by the proposed nonlocal dielectric functions have been found in the
approximation of specular reflection. This made it possible to calculate the
effective Casimir pressure between two parallel metallic plates, the Casimir force
between a sphere and a plate, and its gradient predicted by the suggested approach
in configurations of several precise experiments performed during the last 15
years by two different experimental groups by means of micromechanical torsional
oscillator and atomic force microscope. In doing so the experiments with both
nonmagnetic and magnetic test bodies were considered.

It was shown that the suggested spatially nonlocal dielectric functions taking into
account the dissipation properties of conduction electrons bring the Lifshitz theory
to equally good agreement with the measurement data of all the performed experiments
as does the plasma model which disregards the dissipation of conduction electrons.
Good agreement with seven considered experiments, which were performed between both
nonmagnetic and magnetic metallic surfaces (Au-Au, Au-Ni, and Ni-Ni), was reached
with the velocity parameters common to the transverse and longitudinal
permittivities $\vTn=\vLn=v_n=3v_{{\rm F},n}/2$, where the Fermi velocities
$v_{{\rm F},n}$ are determined in the approximation of a spherical Fermi surface.
In doing so,
the theoretical predictions are rather sensitive to the value
of $\vTn$ but are almost
independent on $\vLn$ in the region from 0 to $10v_{{\rm F},n}$.
In the regions of experimental separations these predictions differ from the previously
made experimentally consistent predictions using the dissipationless plasma models
only in the limits of measurement errors.

The single precise experiment which was not compared with theoretical predictions
of the suggested nonlocal approach to calculation of the Casimir force is the
measurement of differential Casimir force between the Ni-Ni surfaces of a sphere
and a plate performed by means of a micromechanical torsional oscillator \cite{33}.
Taking into account, however, that experiments by means of both an atomic force
microscope and micromechanical torsional oscillator are in a good agreement with
this calculation approach for Au surfaces, the results of Sec.~VI for two magnetic
surfaces can be safely extended to the experiment of Ref.~\cite{33}.

To conclude, the suggested spatially nonlocal dielectric functions
include the dissipation of conduction electrons
leading to only negligibly small deviations from the
standard Drude dielectric response in the area of propagating waves on the mass
shell,
satisfy the Kramers-Kronig relations,
and bring the theoretical predictions of the Lifshitz theory in agreement
with the measurement data of all precise experiments on measuring the Casimir force.
In the future it is desirable to provide some grounding in theory to these
permittivities based, e.g., on the polarization tensor in (3+1)-dimensions
\cite{105},  or quantum field theoretical approach to correlation functions and
Boltzmann kinetic theory \cite{106}.

\section*{ACKNOWLEDGMENTS}

This work was partially
supported by the Peter the Great Saint Petersburg Polytechnic
University in the framework of the Russian state assignment for basic research
(project No.\ FSEG-2020-0024).
This paper has been supported by the  Kazan Federal University
Strategic Academic Leadership Program.
The authors are grateful to C.\ Henkel and V.\ B.\ Svetovoy for useful
discussions.


\end{document}